\documentclass[twocolumn]{aastex701}
\usepackage{amsmath}

\submitjournal{ApJL}
\shorttitle{A Late Rise in Tidal Dissipation}
\shortauthors{Jiang}

\begin{document}

\title{Old World, New Fire: A Late Rise in L 98-59 b's Tidal Dissipation}

\correspondingauthor{Qunfeng Jiang}
\author[0009-0008-0600-3743]{Qunfeng Jiang}
\affiliation{Independent Researcher, Shanghai, China}
\email[show]{qfjiang01@gmail.com}

\begin{abstract}
Atmospheric observations and interior calculations identify the $4.94$~Gyr-old sub-Earth L~98-59~b as a candidate volcanically active world.  An SO$_2$ atmosphere would require rapid replenishment, while coupled interior--atmosphere models permit a present tidal heat flux near $52\ \mathrm{W\,m^{-2}}$.  However, the dissipation that supplies this heat also damps the eccentricity that powers it, raising the question of how long the present high-heat state can survive.  We draw correlated orbital configurations from the five-planet radial-velocity--transit-timing posterior and evolve their secular modes backward with relativistic precession and tides in b.  For low-to-moderate early eccentricities, mean fluxes of $52$ and $65\ \mathrm{W\,m^{-2}}$ have maximum lookback times of $129^{+118}_{-56}$ and $103^{+95}_{-45}$~Myr, respectively; every sampled orbit exhausts its budget well before the system age.  A constant dissipation efficiency over $4.94$~Gyr permits a median present flux of only $1.36\ \mathrm{W\,m^{-2}}$.  If the strong-dissipation state began $50$~Myr ago, half of the orbital posterior requires early dissipation to be weaker by factors of at least 62 and 92.  A late rheological transition therefore offers a coherent route to young tidal fire on an old world.
\end{abstract}


\section{Introduction}\label{sec:intro}

L~98-59 is a nearby M-dwarf system whose compact architecture joins atmospheric, orbital, and interior measurements in a single laboratory.  Three transiting terrestrial planets orbit in 2.25, 3.69, and 7.45~days, and radial velocities reveal two additional planets on 12.83 and 23.06~day orbits \citep{Kostov2019,Cadieux2025}.  The host is $4.94\pm0.28$~Gyr old \citep{EngleGuinan2023,Cadieux2025}.  The innermost planet b has $R_b=0.837\pm0.019\,R_\oplus$ and $M_b=0.46\pm0.11\,M_\oplus$.  Its small orbit and measured nonzero eccentricity make L~98-59 the most favorable known multiplanet system for strong tidal melting in a comparative survey of rocky exoplanets \citep{Seligman2024}.

JWST/NIRSpec observed four transits of L~98-59~b.  Their combined transmission spectrum remains consistent with a bare rock, while atmospheric retrievals give a reported $3.6\sigma$ Bayesian preference for an SO$_2$-rich atmosphere over a featureless spectrum \citep{BelloArufe2025}.  The atmospheric interpretation requires rapid replenishment.  For a present XUV flux one tenth of the stellar saturation value and a 1\% escape efficiency, the energy-limited loss rate is approximately $2\times10^5\ \mathrm{kg\,s^{-1}}$; an unreplenished 10~bar atmosphere then lasts only $\sim10$~Myr.  Balancing that loss with terrestrial volatile abundances in the erupted material requires an extrusive rate of $1$--$2\times10^9\ \mathrm{kg\,s^{-1}}$, or at least eight times Io's volcanic and tidal power per unit planetary mass \citep{BelloArufe2025}.  The atmosphere can therefore act as a calorimeter that integrates interior power over many eruption cycles.

Eccentricity tides both generate heat and consume the eccentricity that powers it: the heat rate scales as $e^2/Q'$, while eccentricity damping scales as $1/Q'$ \citep{Peale1979,Segatz1988}.  Stronger dissipation therefore raises the tidal heat flux but shortens the time over which a given eccentricity can sustain that flux.  Planetary companions can extend the heating by maintaining a forced eccentricity.  In a compact system, secular interactions distribute eccentricity among collective modes, allowing outer planets to feed a dissipating inner orbit \citep{Mardling2007,GreenbergVanLaerhoven2011,HansenMurray2015}.  The available angular-momentum deficit then acts as a finite orbital reservoir whose depletion limits the duration of the heating \citep{FarhatChiang2026}.  This balance between forcing and damping shapes the tidal histories of systems such as the resonant TRAPPIST-1 chain and LP~791-18 \citep{Barr2018,Peterson2023}; for LP~791-18~d, imposing a tidally damped free eccentricity substantially changes the inferred heating \citep{GreklekMcKeon2025}.

Previous studies place L~98-59~b in a strong tidal-heating regime.  The coupled early-evolution models of \citet{Nicholls2025} give a median tidal flux of $52\ \mathrm{W\,m^{-2}}$ for b at global energy balance.  The minimum volcanic power needed to replenish the candidate SO$_2$ atmosphere corresponds to about $65\ \mathrm{W\,m^{-2}}$ when scaled to the mass and radius of b \citep{BelloArufe2025}.  The dissipation needed to produce either flux also damps the eccentricity that powers it.  This tension raises a question: \textit{How long could L~98-59~b have maintained the tidal dissipation efficiency required to produce either flux today?}

We turn that tension into a tidal clock using the correlated five-planet orbital posterior.  Section~\ref{sec:methods} describes the secular-mode calculation, heat-flux mapping, and two-stage dissipation model.  Section~\ref{sec:results} presents the clock and the inferred contrast between early and present dissipation.  Section~\ref{sec:discussion} examines atmospheric averaging, a possible c--d 2:1 excitation event, and the interior evolution implied by the clock; Section~\ref{sec:conclusions} summarizes the resulting history of L~98-59~b.

\section{Methods}\label{sec:methods}

\subsection{Correlated orbital samples and heat-flux estimates}\label{subsec:posterior}

We use the fixed-inclination joint radial-velocity (RV)--transit-timing-variation (TTV) posterior of \citet{Cadieux2025}.  A posterior sample contains the stellar mass and the five planets' mutually correlated RV semiamplitudes and eccentricity coordinates.  The orbital periods and epochs of inferior conjunction of planets e and f are also sampled.  For planets b--d, the linear ephemerides ($P$ and $t_0$) are held fixed at the values determined from the transit data.  We analyze 5000 randomly selected complete samples, retaining the mass--eccentricity, c--d, and apsidal correlations within each draw.

The fitted eccentricity coordinates are
\begin{equation}
h_i=\sqrt{e_i}\cos\varpi_i,\qquad
k_i=\sqrt{e_i}\sin\varpi_i,
\label{eq:hk}
\end{equation}
so that $e_i=h_i^2+k_i^2$ and $\varpi_i=\mathrm{atan2}(k_i,h_i)$.  Here $\varpi_i$ is the longitude of periapse in the coplanar secular model.  We obtain each planet mass by numerically inverting
\begin{equation}
K_i=\left(\frac{2\pi G}{P_i}\right)^{1/3}
\frac{M_i\sin i_i}{(M_\star+M_i)^{2/3}\sqrt{1-e_i^2}},
\label{eq:rv_mass}
\end{equation}
using the same Keplerian RV convention and fixed inclinations as \citet{Cadieux2025}, and compute $a_i$ from Kepler's third law.  A Gaussian draw from $0.837\pm0.019\,R_\oplus$ supplies the radius of b.  The resulting 16th, 50th, and 84th percentiles reproduce the masses and eccentricities in Table~C.1 of \citet{Cadieux2025}; in particular, $e_b=0.031^{+0.017}_{-0.016}$.

For a synchronously rotating planet with small eccentricity and negligible obliquity, the constant-phase-lag heating rate gives the surface flux \citep{Peale1979,Segatz1988}
\begin{equation}
F_{\rm tide}=\frac{63}{16\pi Q'_b}
\frac{G M_\star^2R_b^3}{a_b^6}n_b\langle e_b^2\rangle .
\label{eq:flux}
\end{equation}
The mean motion is $n_b=2\pi/P_b$.  The conventional quality factor $Q$ is $2\pi$ times the maximum energy stored in the tidal deformation divided by the energy dissipated per cycle; $k_2$ is the degree-two potential Love number.  We combine them as $Q'=3Q/(2k_2)$.  Thus $1/Q'$ measures the effective dissipation per unit forcing without assigning a specific interior structure.

The same convention gives the eccentricity-amplitude damping rate, where $\tau_{e,b}$ is the e-folding time of $e_b$ for an otherwise isolated b:
\begin{equation}
\tau_{e,b}^{-1}=\frac{63}{4Q'_b}\frac{M_\star}{M_b}
\left(\frac{R_b}{a_b}\right)^5 n_b .
\label{eq:damping}
\end{equation}
Equations~(\ref{eq:flux}) and (\ref{eq:damping}) use one $Q'_b$, so the heat produced and the orbital reservoir consumed are energetically linked.  Tides in c--f are set to zero in the clock calculation.  This isolates the cost of the proposed volcanic state on b and keeps the exterior secular reservoir available as long as possible.

We perform separate calculations using two heat-flux estimates from the literature.  We adopt $52\ \mathrm{W\,m^{-2}}$ from the median flux for b at global energy balance in the coupled interior--atmosphere models of \citet{Nicholls2025}.  The second estimate follows from the atmospheric replenishment calculation of \citet{BelloArufe2025}, which requires at least eight times Io's heating per unit mass.  Converting the astrometric Io flux $F_{\rm Io}=2.24\pm0.45\ \mathrm{W\,m^{-2}}$ \citep{Lainey2009} gives
\begin{equation}
F_b\simeq 8F_{\rm Io}\frac{M_b}{M_{\rm Io}}
\left(\frac{R_{\rm Io}}{R_b}\right)^2\simeq65\ \mathrm{W\,m^{-2}}.
\label{eq:io_scaling}
\end{equation}
This conversion maps the unit-mass heating requirement of \citet{BelloArufe2025} to a surface flux; its uncertainty is dominated by the atmospheric escape and volatile-supply assumptions.

\subsection{Five-planet secular modes and the tidal clock}\label{subsec:clock}

Secular dynamics are most compactly written with the complex eccentricity $z_i=e_i\exp(i\varpi_i)$.  To second order in eccentricity and first order in planet mass, the five-component vector $\mathbf z=(z_b,\ldots,z_f)^\mathsf{T}$ obeys Laplace--Lagrange theory \citep{MurrayDermott1999}:
\begin{equation}
\dot{\mathbf z}=i\mathbf A\mathbf z.
\label{eq:ll}
\end{equation}
For planet $i$ perturbed by $j$, the matrix elements are
\begin{align}
A_{ii}^{\rm pp}&=\frac{n_i}{4}\sum_{j\ne i}
\frac{M_j}{M_\star+M_i}\alpha_{ij}\bar\alpha_{ij}
b_{3/2}^{(1)}(\alpha_{ij}),\nonumber\\
A_{ij}^{\rm pp}&=-\frac{n_i}{4}\frac{M_j}{M_\star+M_i}
\alpha_{ij}\bar\alpha_{ij}b_{3/2}^{(2)}(\alpha_{ij}),
\label{eq:matrix}
\end{align}
where $\alpha_{ij}=\min(a_i,a_j)/\max(a_i,a_j)$, $\bar\alpha_{ij}=\alpha_{ij}$ when $i$ is interior to $j$ and unity otherwise, and $b_s^{(m)}$ is the Laplace coefficient.  General relativity adds the diagonal precession
\begin{equation}
A_{ii}^{\rm GR}=\frac{3GM_\star n_i}{a_i c^2},
\label{eq:gr}
\end{equation}
so $\mathbf A=\mathbf A^{\rm pp}+\operatorname{diag}(A_{ii}^{\rm GR})$.  Relativistic precession matters most for b and changes how efficiently the outer modes project onto its eccentricity.

Tidal damping makes the secular operator non-Hermitian:
\begin{equation}
\dot{\mathbf z}=\left(i\mathbf A-\mathbf\Gamma\right)\mathbf z,
\qquad
\mathbf\Gamma=\operatorname{diag}(\tau_{e,b}^{-1},0,0,0,0).
\label{eq:damped_operator}
\end{equation}
We diagonalize this operator for each posterior sample and each trial $Q'_b$.  Its five eigenvectors $\mathbf V_j$ describe coherent apsidal patterns across b--f; the complex eigenvalues contain the precession frequency $g_j$ and the positive amplitude-damping rate $\gamma_j$.  Projecting the observed state onto the eigenbasis gives
\begin{equation}
z_i(t)=\sum_{j=1}^{5} V_{ij}C_j
\exp[(ig_j-\gamma_j)t].
\label{eq:modes}
\end{equation}
The coefficient $C_j$ contains the amplitude and phase of mode $j$ in that correlated posterior sample.  Averaging over the incommensurate secular phases removes cross terms and yields
\begin{equation}
\langle e_i^2\rangle(t)=\sum_{j=1}^{5}|V_{ij}C_j|^2e^{-2\gamma_jt}.
\label{eq:rms}
\end{equation}
This is the secular root-mean-square (rms) eccentricity: $e_{i,\rm RMS}=\sqrt{\langle e_i^2\rangle}$.  It is distinct from the posterior median of the osculating $e_i$ and is the quantity entering the long-term mean heating.

For a chosen present flux, we iterate Equation~(\ref{eq:flux}) with the damped eigenvectors until the inferred $Q'_b$ and $\langle e_b^2\rangle$ are mutually consistent.  We then reverse Equation~(\ref{eq:rms}); a damped mode grows as $e^{+\gamma_jt}$ at lookback time $t$.  We define $t_{\rm clock}$ as the first lookback time at which any planet reaches an adopted rms eccentricity limit $e_{\rm lim}$.  The baseline $e_{\rm lim}=0.1$ is 2.4 times the median present $e_{b,\rm RMS}$ and several times the characteristic few-percent eccentricity of compact Kepler multiplanet systems \citep{VanEylenAlbrecht2015}, while remaining within the low-eccentricity regime of the secular expansion.  We also calculate $e_{\rm lim}=0.2$.  Because angular-momentum deficit scales as $e^2$ at small eccentricity, doubling the limit allows four times the eccentricity-squared budget for a given orbital component.  A common excitation near 0.2 would also bring several adjacent radial excursions close to overlap and place fourth-order secular terms beyond the accuracy of the linear model.  We therefore use 0.1 as the baseline and 0.2 as a deliberately permissive sensitivity bound.

The backward calculation answers how long the present or stronger dissipation state can be traced before it requires a highly excited architecture.  We also invert it.  For every posterior sample, a root finder determines $Q'_{b,\rm crit}$ such that $t_{\rm clock}=T_\star=4.94$~Gyr, recomputing the eigenvectors and damping rates at every step.  Substitution into Equation~(\ref{eq:flux}) gives $F_{\rm crit}$, the largest present heat flux compatible with one constant dissipation efficiency over the system age.

The tidal clock relies on linear secular theory, so we test its eccentricity amplitudes against direct five-planet integrations.  We select 30 posterior samples spanning the full $52\ \mathrm{W\,m^{-2}}$ clock distribution and the high- and low-eccentricity tails of c--d and f.  Each N-body system uses the same masses and osculating orbital elements as its secular calculation at BJD$-2457000=1345$, includes mutual gravity and first-order relativistic precession, and is integrated without imposed migration or tidal damping for 8000~yr using REBOUND/WHFast with a step $P_b/20$ \citep{ReinLiu2012,ReinTamayo2015,Tamayo2020}.  This interval covers several cycles of every relevant secular mode.  We measure $e_{b,\rm RMS}$ directly from each time series and compare it with the linear prediction for the same initial state.  All 30 systems remain stable; the N-body-to-linear amplitude ratio has 16th, 50th, and 84th percentiles of 0.865, 0.906, and 0.980, and the largest difference is 20.7\%.  The posterior clock uses the transparent linear solution, and this comparison sets a 9\% typical and 21\% maximum scale for nonlinear amplitude corrections.

\subsection{Atmospheric averaging and a two-stage dissipation history}\label{subsec:twostage}

The candidate atmosphere responds to the integrated volcanic supply.  We represent that response by an exponential reservoir,
\begin{equation}
\bar F_{\rm atm}(t)=\frac{1}{\tau_{\rm atm}}
\int_{-\infty}^{t}F_{\rm tide}(t')
e^{-(t-t')/\tau_{\rm atm}}\,dt',
\label{eq:atmosphere_filter}
\end{equation}
with $\tau_{\rm atm}=10$~Myr from the 10~bar escape estimate of \citet{BelloArufe2025}.  We apply this filter to a representative 8000~yr N-body heat-flux time series periodically continued to steady state.  Response times of 1 and 100~Myr show how the result scales around the baseline.  Because every response time greatly exceeds the secular periods, Equation~(\ref{eq:atmosphere_filter}) approaches the $\langle e_b^2\rangle$ average used in Equation~(\ref{eq:flux}).

To place the clock in a 4.94~Gyr history, we assign the most recent interval $t_{\rm recent}$ to the present efficiency $Q'_{\rm now}$ and the preceding interval to $Q'_{\rm early}$.  In the weak-damping limit all modal damping rates scale as $1/Q'$.  The elapsed history expressed in units of present-day dissipation is therefore
\begin{equation}
t_{\rm eq}=t_{\rm recent}+(T_\star-t_{\rm recent})
\frac{Q'_{\rm now}}{Q'_{\rm early}}.
\label{eq:budget}
\end{equation}
Compatibility requires $t_{\rm eq}\leq t_{\rm clock}$.  We solve this inequality with each sample's paired $Q'_{\rm now}$ and $t_{\rm clock}$.  If $t_{\rm recent}>t_{\rm clock}$, the sample remains incompatible even as $Q'_{\rm early}\rightarrow\infty$ and stays in the denominator of the posterior fraction.  The reported curve is the minimum $Q'_{\rm early}$ for which 50\% of all samples are compatible.

Equation~(\ref{eq:budget}) also covers a continuous history after replacing $t_{\rm eq}$ by
\begin{equation}
t_{\rm eq}=\int_0^{T_\star}D(t)\,dt,\qquad
D(t)\equiv\frac{1/Q'(t)}{1/Q'_{\rm now}}
=\frac{Q'_{\rm now}}{Q'(t)}.
\label{eq:dissipation_integral}
\end{equation}
The dimensionless $D(t)$ is the dissipation strength relative to today.  Its time integral is the orbital cost of an arbitrary rheological transition.

\section{Results}\label{sec:results}

\subsection{The high-heat state has a short dynamical lifetime}\label{subsec:clockresults}

\begin{figure*}[t]
\centering
\includegraphics[width=0.96\textwidth]{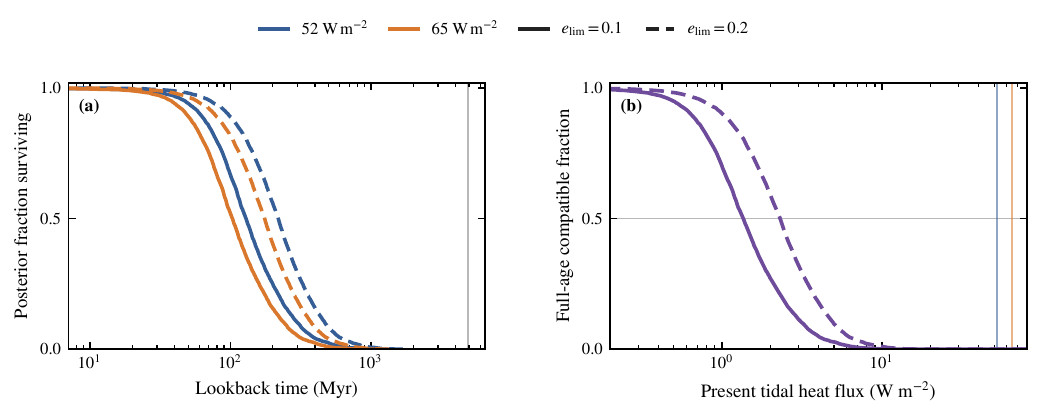}
\caption{Two views of the finite eccentricity reservoir.  (a) Fraction of the joint orbital posterior whose present dissipation efficiency can be propagated backward for at least the plotted time.  Colors set the present secularly averaged heat flux; solid and dashed curves adopt $e_{\rm lim}=0.1$ and 0.2.  The system age lies at the right edge.  (b) Fraction compatible with a constant dissipation efficiency for the full 4.94~Gyr as a function of present heat flux.  The high-flux atmospheric and interior estimates lie beyond the displayed posterior support.  Numerical summaries appear in Table~\ref{tab:clocks}.}\label{fig:clockflux}
\end{figure*}

The posterior secular rms eccentricity of b is $0.0415^{+0.0151}_{-0.0137}$.  Producing $52\ \mathrm{W\,m^{-2}}$ requires $Q'_b=3293^{+2836}_{-1815}$; producing $65\ \mathrm{W\,m^{-2}}$ requires $Q'_b=2634^{+2269}_{-1452}$.  These values convert the current architecture into short direct damping times and, after interplanetary mode exchange is included, the clock distributions in Figure~\ref{fig:clockflux}a and Table~\ref{tab:clocks}.

\begin{deluxetable*}{lcccc}
\tablecaption{Present Dissipation and Posterior Tidal Clocks\label{tab:clocks}}
\tablehead{
\colhead{$F_{\rm now}$} &
\colhead{$Q'_b$} &
\colhead{$\tau_{e,b}$} &
\colhead{$t_{\rm clock}$, $e_{\rm lim}=0.1$} &
\colhead{$t_{\rm clock}$, $e_{\rm lim}=0.2$}\\
\colhead{$(\mathrm{W\,m^{-2}})$} & \colhead{} & \colhead{(Myr)} & \colhead{(Myr)} & \colhead{(Myr)}}
\startdata
52 & $3293^{+2836}_{-1815}$ & $91^{+74}_{-49}$ & $129^{+118}_{-56}$ & $219^{+179}_{-101}$\\
65 & $2634^{+2269}_{-1452}$ & $73^{+59}_{-39}$ & $103^{+95}_{-45}$ & $175^{+143}_{-81}$
\enddata
\tablecomments{Reported intervals are the 16th--84th posterior percentiles.  $\tau_{e,b}$ is the isolated eccentricity-amplitude e-folding time from Equation~(\ref{eq:damping}); $t_{\rm clock}$ includes all five secular modes.}
\end{deluxetable*}

For $e_{\rm lim}=0.1$, the median clocks are 129 and 103~Myr.  Ninety-five percent of the posterior has exhausted the corresponding budgets by 379 and 303~Myr.  The longest individual clocks are 1.31 and 1.05~Gyr, still far below the system age.  In every sample, b is the first planet to reach the adopted rms limit, so the b-weighted fast modes trigger the clock throughout the posterior.

The $e_{\rm lim}=0.2$ calculation expands the available early angular-momentum deficit.  The medians increase to 219 and 175~Myr, while 95\% of samples are exhausted by 584 and 467~Myr.  Even the longest posterior examples end by 1.82 and 1.46~Gyr.  Permitting a much warmer early system lengthens the clock by less than a factor of two because the backward growth is exponential and dominated by the same damped modes.

Figure~\ref{fig:clockflux}b removes the choice of a particular high-flux estimate.  The present flux that can accompany a constant 4.94~Gyr efficiency is
\begin{equation}
F_{\rm crit}=1.36^{+1.24}_{-0.59}\ \mathrm{W\,m^{-2}}
\end{equation}
for $e_{\rm lim}=0.1$, and $2.30^{+1.88}_{-1.07}\ \mathrm{W\,m^{-2}}$ for $e_{\rm lim}=0.2$.  No sample under either limit supports $52$ or $65\ \mathrm{W\,m^{-2}}$ for the system age.  The critical-flux distribution provides a posterior-based dividing line: heat of order $1\ \mathrm{W\,m^{-2}}$ can be ancient at constant efficiency, while heat of order tens of $\mathrm{W\,m^{-2}}$ requires a time-dependent dissipation history or recent orbital excitation.

At $52\ \mathrm{W\,m^{-2}}$, the longest-lived mode has a median amplitude lifetime of 15.5~Gyr, yet contributes only 6.4\% of b's present $e^2$ and places only 0.6\% of its angular-momentum deficit in b.  The mode contributing the largest share of b's $e^2$ has a median lifetime of 197~Myr, contributes 52.9\% of b's present eccentricity power, and typically places 77\% of its angular-momentum deficit in b.

\subsection{Constraints on weaker early dissipation}\label{subsec:twostageresults}

\begin{figure*}[t]
\centering
\includegraphics[width=0.93\textwidth]{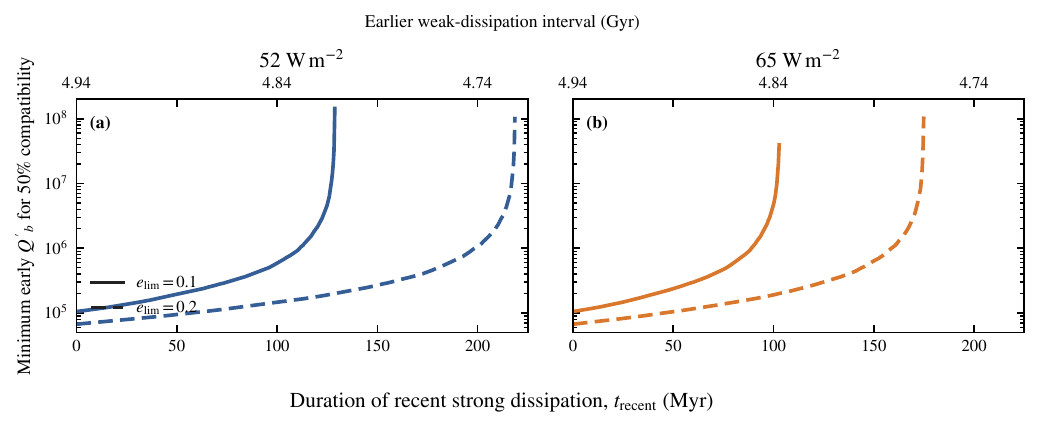}
\caption{Minimum early $Q'_b$ required for 50\% of the complete orbital posterior to survive for 4.94~Gyr.  Panels show the $52$ and $65\ \mathrm{W\,m^{-2}}$ present heat-flux cases; solid and dashed curves use the two early eccentricity limits.  The lower horizontal axis gives the duration of the recent strong-dissipation state and the upper axis gives the preceding weak-dissipation interval.  Larger $Q'_{\rm early}$ means weaker early dissipation.  A curve ends when the recent interval alone has consumed the median orbital budget.}\label{fig:twostage}
\end{figure*}

Figure~\ref{fig:twostage} distributes the same orbital budget between an early and a recent phase.  For a $50$~Myr present-strength interval and $e_{\rm lim}=0.1$, half of the posterior requires
\begin{align}
Q'_{\rm early}&\geq1.95\times10^5 &&(52\ \mathrm{W\,m^{-2}}),\nonumber\\
Q'_{\rm early}&\geq2.37\times10^5 &&(65\ \mathrm{W\,m^{-2}}).
\end{align}
Relative to the paired present values, these are dissipation strengths smaller by factors of 62 and 92.  In the limit of negligible early dissipation, 94.0\% and 89.1\% of samples can still accommodate a 50~Myr recent interval.  The remaining samples possess clocks shorter than 50~Myr, so the recent phase alone already exceeds their budgets.

The broader $e_{\rm lim}=0.2$ reservoir lowers the required contrasts at 50~Myr to factors of 29 and 39.  In all cases the curves steepen as $t_{\rm recent}$ approaches the median clock, because the remaining early budget $t_{\rm clock}-t_{\rm recent}$ tends to zero.  Each endpoint marks the longest possible present-strength episode for the corresponding posterior median, including the limit of vanishing early tides.

\section{Discussion}\label{sec:discussion}

\subsection{The atmosphere measures the secular mean}\label{subsec:atmosphere}

The osculating eccentricity of b varies rapidly, so an instantaneous tidal flux can look far more extreme than the sustained volcanic supply.  In the representative five-planet N-body time series, scaled to the $52\ \mathrm{W\,m^{-2}}$ calculation, the raw flux ranges from $1.8\times10^{-3}$ to $226\ \mathrm{W\,m^{-2}}$ over secular cycles.  Its median is $36.5\ \mathrm{W\,m^{-2}}$ and its time mean is $51.18\ \mathrm{W\,m^{-2}}$.  The skew arises from $F\propto e^2$: short high-eccentricity intervals carry substantial energy without representing a separate long-lived state.

Equation~(\ref{eq:atmosphere_filter}) suppresses this rapid instantaneous variability.  A 1~Myr response leaves only $0.024\ \mathrm{W\,m^{-2}}$ peak-to-peak variation.  The 10~Myr response leaves $0.0024\ \mathrm{W\,m^{-2}}$, or $4.6\times10^{-5}$ of the mean, and a 100~Myr response leaves $2.4\times10^{-4}\ \mathrm{W\,m^{-2}}$.  The atmospheric-loss time spans thousands of secular cycles, so the atmosphere effectively sees $\langle e_b^2\rangle$.  A brief $226\ \mathrm{W\,m^{-2}}$ maximum contributes to the mean yet supplies too little integrated energy on its own; the amplitude of the entire secular mode must remain large over Myr timescales.

This comparison also clarifies the meaning of the clock.  The 103--129~Myr medians are more than ten times the estimated 10~Myr lifetime of a 10~bar SO$_2$ atmosphere.  A recent high-dissipation phase can therefore reach atmospheric steady state through repeated replenishment.  Its duration remains only a few percent of the stellar age.  The atmospheric observation and the orbit can be reconciled when the current supply is persistent on atmospheric timescales and young on planetary timescales.

\subsection{A c--d 2:1 passage fails the timing test}\label{subsec:resonance}

The c--d pair lies just wide of the 2:1 commensurability, with $P_d/P_c=2.0188$ \citep{Cadieux2025}.  A recent divergent passage across 2:1 could inject free eccentricity into the five-planet secular modes.  We test the strongest version of that idea by prescribing inward migration of c from $P_d/P_c=1.995$ to the observed ratio.  This experiment asks how much mode amplitude a smooth crossing can deliver and whether the same tide can plausibly move the pair to its current separation.

We select three joint posterior samples at the 16th, 50th, and 84th percentiles of the fast-mode amplitude required 50~Myr ago.  For each sample, d--f are matched to the present longest-lived mode and projected onto the corresponding pre-crossing eigenvector, leaving a pure slow initial secular state.  Planet c is shifted outward to set the initial period ratio; all five masses are retained.  We sample the unknown conjunction geometry by setting the 2:1 resonant angle
\begin{equation}
\phi_c=2\lambda_d-\lambda_c-\varpi_c
\end{equation}
to six evenly spaced values, $\phi_c=j\pi/3$ for $j=0,\ldots,5$.  We impose migration times of 100, 500, and 2000~yr, plus 5000~yr for the median posterior sample, stop migration at the observed period ratio, and continue for 20~yr.  The WHFast step is $P_b/10$; two representative calculations repeated at $P_b/20$ retain their classifications.  The resulting grid contains 60 independent physical trials and two time-step convergence reruns.  We assess adiabaticity using the resonant timescales.  In the 14 of 18 independent 2000~yr trials that possess a local island at nominal 2:1, its median period-ratio width is 0.0115, the imposed drift crosses that width in 970~yr, and the local libration period is 5.58~yr.  The median ratio of these timescales is 173 (16th--84th percentiles 160--180), placing these passages in the locally adiabatic regime \citep{MustillWyatt2011}; the other four have $\Gamma\leq3/2$ and no local island at the nominal crossing.

The imposed crossing satisfies the excitation requirement alone.  For the 2000~yr sweeps, the injected fast-mode rms amplitude is $0.059^{+0.014}_{-0.009}$.  After damping at the paired $Q'_b$, this supports the present $52\ \mathrm{W\,m^{-2}}$ state for a median of 53~Myr and the $65\ \mathrm{W\,m^{-2}}$ state for 42~Myr.  The 500--5000~yr sweeps converge to amplitudes of approximately 0.05--0.07, showing that the immediate dynamical response is large enough to test the subsequent timing requirement.  Appendix~\ref{app:resonance} applies a resonant-island phase-space criterion and identifies measurable perturbations among securely exterior passages as well as trajectories initialized inside the island; capture is therefore unnecessary for a transient excitation.  However, c tides cannot move the pair from 2:1 to its present separation within the lifetime of the injected b mode.  At fixed orbital angular momentum, reaching $P_d/P_c=2.0188$ while ending at the observed small $e_c$ requires $e_c\simeq0.079$ at resonance.  Completing that evolution in 100~Myr requires $Q'_c=255^{+68}_{-58}$; 50 and 200~Myr require median values of 127 and 510.  For $Q'_c=3300$, comparable to b in the $52\ \mathrm{W\,m^{-2}}$ state, the median travel time is 1.30~Gyr, far longer than the 53 and 42~Myr over which the injected mode can support the two high-heat states.  Accelerated N-body circularization tests give the same conflict: trajectories starting at $e_c=0.03$--0.06 never reach the observed period ratio, while $e_c=0.08$ succeeds only for a restricted initial phase.

Thus, although passage through the c--d 2:1 commensurability can momentarily excite b's fast secular mode, the required post-crossing evolution erases that mode before the system reaches its present architecture.  The tested c-tide-only passage cannot account for today's high tidal heat flux.  A late rise in b's dissipation remains the history that satisfies both the old planetary system and the young volcanic state.

\subsection{A late rheological transition: an old world with young fire}\label{subsec:rheology}

At either heat flux, our secular calculations show that sustained strong dissipation would deplete the eccentricity that powers b's tides long before the system reaches its current age, even with forcing by its companions.  This contrast between an old system and a short-lived high-heat state can be explained by a two-stage history in which weak early dissipation preserves the eccentricity reservoir and a late increase in dissipation produces strong heating today.

Mantle cooling and partial crystallization provide a mechanism for a late increase in dissipation.  As the mantle cools, changes in melt fraction and in the thickness and depth of partially molten layers alter the tidal response.  Models of Io show that partially molten layers can dominate tidal dissipation, with the total dissipated power depending strongly on the assumed rheology, melt fraction, and interior structure \citep{Kervazo2021}.  Coupled models of rocky exoplanets also show that tidal dissipation can peak as the mantle cools through a temperature range where its viscoelastic response is strongest \citep{DriscollBarnes2015}.  A shift into this regime may give L~98-59~b a physical route from weak early dissipation to the stronger dissipation in our two-stage model.

The coupled mantle--atmosphere models of \citet{Nicholls2025} illustrate how L~98-59~b could maintain a partially molten mantle: as the mantle cooled toward its critical melt fraction, tidal heating increased, while the outgassed atmosphere slowed radiative cooling.  At their reference eccentricity of $e_b=0.167$, b reached global energy balance within 10~Myr, with a median mantle melt fraction of 38.3\% and a median tidal flux of $52\ \mathrm{W\,m^{-2}}$.  Their calculations therefore establish a physically plausible strongly heated state.  However, they assume orbital steady state: the eccentricity is fixed within each calculation, and neither secular exchange among the five planets nor tidal depletion of their shared eccentricity reservoir is evolved.  Their reference eccentricity is also substantially larger than the updated joint RV--TTV estimate, $e_b=0.031^{+0.017}_{-0.016}$ \citep{Cadieux2025}.  These differences directly affect the lifetime of the heated state, because producing the same mean flux at a smaller eccentricity requires stronger dissipation and therefore faster damping.  Existing models therefore show that tidal dissipation can change substantially during thermal evolution, but they do not yet explain why L~98-59~b would enter a high-dissipation state only recently.  Testing that timing will require a coupled thermal--orbital evolution model that includes atmospheric insulation and escape.

\begin{figure*}[t]
\centering
\includegraphics[width=0.76\textwidth]{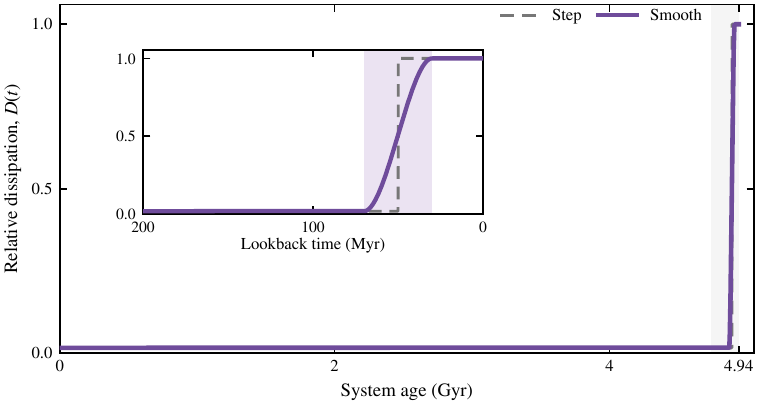}
\caption{Illustrative two-stage histories.  Dissipation rises smoothly from 70 to 30~Myr ago (solid) or steps up 50~Myr ago (dashed).  Both curves give a present-equivalent dissipation time of $\int D\,dt=131$~Myr.  The inset shows the final 200~Myr.}\label{fig:dissipationhistory}
\end{figure*}

For the illustrative case in which the present strong state began 50~Myr ago, the median $52\ \mathrm{W\,m^{-2}}$ history requires $Q'_{\rm early}\simeq2\times10^5$ and $Q'_{\rm now}\simeq3300$, so the relative dissipation rises from $D\simeq0.016$ to unity.  Figure~\ref{fig:dissipationhistory} compares this step with a smooth rise from 70 to 30~Myr ago that gives the same cumulative dissipation.  The two-stage model shows how a late increase in dissipation could reconcile the present heating requirement with the system's orbital history.  A self-consistent calculation over the system's lifetime must follow the evolution of the five orbits together with mantle cooling, crystallization, and atmospheric outgassing and escape.  It must also account for dissipation in the solid mantle and in any partially molten layers or magma ocean \citep{FarhatChiang2026}.

JWST emission spectra of 55~Cancri~e favor a volatile secondary atmosphere that may be sustained by exchange with a magma ocean \citep{Hu2024}.  LP~791-18~d and the TRAPPIST-1 planets illustrate how neighboring planets can maintain the eccentricity that powers tidal heating \citep{Peterson2023,Barr2018}.  Together, these systems provide complementary access to a planet's tidal history through the observable consequences of interior heat and the orbital architecture that sustains eccentricity.  Combining heat-flux estimates, system ages, and correlated orbital constraints could help constrain the coupled thermal--orbital histories of tidally active rocky planets from present-day observations.

\subsection{Scope and observational tests}\label{subsec:limits}

Three approximations set the present precision.  First, a constant $Q'$ within each phase compresses a frequency- and temperature-dependent complex Love number into one effective parameter.  This is the appropriate quantity for the orbital energy budget, while a thermal calculation is needed to map it onto viscosity, melt topology, and composition.  Second, linear secular theory neglects higher-order mode coupling and semimajor-axis evolution.  The 30 posterior N-body integrations show a median 9.4\% lower $e_{b,\rm RMS}$ and at most a 20.7\% amplitude difference.  Those corrections are material for a final coupled fit and much smaller than the factor of 38 between the $52\ \mathrm{W\,m^{-2}}$ heat-flux case and the median full-age flux limit.  The $e_{\rm lim}=0.2$ results bracket a larger reservoir while marking the edge of the linear, non-crossing regime.

A third source of uncertainty is the present heat flux.  The values of 52 and $65\ \mathrm{W\,m^{-2}}$ come from an interior model and an atmospheric replenishment estimate, respectively.  The $F_{\rm crit}$ distribution makes the dynamical result portable: any future model can compare its present secular-mean tidal flux with the $1.36^{+1.24}_{-0.59}\ \mathrm{W\,m^{-2}}$ constant-history boundary.  Confirmation of an SO$_2$ atmosphere, improved estimates of escape efficiency, and tighter eccentricity constraints would narrow the range of plausible present tidal heat fluxes.  A frequency-dependent mantle--magma-ocean model can then evolve each joint orbital sample with a physical $Q'(T,\phi,\omega)$ and test when dissipation can increase enough to supply the present heating.

Future atmospheric constraints provide a direct test of this history.  If b's atmosphere requires tens of $\mathrm{W\,m^{-2}}$ of sustained tidal supply, the present high-dissipation state occupies only the most recent few percent of the system history.  If future observations favor a heat flux near $1\ \mathrm{W\,m^{-2}}$, an ancient constant-efficiency state remains allowed.  Atmospheric characterization can thus distinguish a long-lived warm world from a brief episode of young fire.

\section{Conclusions}\label{sec:conclusions}

The five-planet architecture supplies a finite forced-eccentricity reservoir.  At the dissipation required for $52$ and $65\ \mathrm{W\,m^{-2}}$, the correlated RV--TTV posterior gives median clocks of $129^{+118}_{-56}$ and $103^{+95}_{-45}$~Myr for a low-to-moderate early architecture.  Ninety-five percent of samples exhaust those budgets within 379 and 303~Myr, and every sample ends before the 4.94~Gyr system age.  Holding the efficiency fixed for the full age restricts the median present flux to $1.36\ \mathrm{W\,m^{-2}}$; a broad early eccentricity budget raises it to $2.30\ \mathrm{W\,m^{-2}}$.

A recent change in effective rheology quantitatively resolves the age contrast.  If the current strong state began 50~Myr ago, half of the posterior requires early dissipation weaker by factors of at least 62 and 92 for the two heat-flux estimates.  A c--d 2:1 passage can excite b's fast mode, although the $\sim1.3$~Gyr c-tide travel time at comparable $Q'$ is too long to preserve that excitation until today in the tested model.  Long-lived early melt followed by late entry into an efficient partially molten dissipation window provides a coherent physical picture.  L~98-59~b can be an old planet whose present atmosphere records a geologically young tidal fire.

\section*{Data and Code Availability}

The analysis code and supporting summary data are available on Zenodo: \dataset[doi:10.5281/zenodo.24502731]{https://doi.org/10.5281/zenodo.24502731}.

\begin{acknowledgments}

The author thanks Juliette Becker for insightful discussions, Charles Cadieux for sharing the joint RV--TTV posterior samples, and Lingshan Xiong for discussions of tidal rheology and interior evolution.  This research has made use of the Astrophysics Data System, funded by NASA under Cooperative Agreement 80NSSC25M7105.  Generative-AI tools were used to assist with code implementation and language editing.  The scientific analysis, interpretation, and conclusions were independently validated by the author, who takes full responsibility for the content of this work.
\end{acknowledgments}

\software{REBOUND v5.1.1 \citep{ReinLiu2012}, WHFast \citep{ReinTamayo2015}, REBOUNDx v5.1.0 \citep{Tamayo2020}, NumPy v2.2.6 \citep{Harris2020}, SciPy v1.15.3 \citep{Virtanen2020}, pandas v2.3.3, Matplotlib v3.10.7 \citep{Hunter2007}}

\appendix
\restartappendixnumbering

\section{Classifying the c--d 2:1 passages}\label{app:resonance}

To determine whether each c--d trajectory enters resonance, we follow its path relative to the instantaneous resonant island.  Near a first-order mean-motion resonance, the two eccentricity-dependent harmonics can be reduced by canonical transformations to a one-degree-of-freedom second fundamental model \citep{HenrardLemaitre1983,MustillWyatt2011,Deck2013}.  We use the two-mass 2:1 coefficients, including the indirect term specific to 2:1, and scale each c--d trajectory into
\begin{equation}
\mathcal H(X,Y;\Gamma)=-\frac{1}{2}
\left[\frac{X^2+Y^2}{2}-\Gamma\right]^2-X.
\label{eq:hamiltonian}
\end{equation}
The canonical action is $J=(X^2+Y^2)/2$ and $X=\sqrt{2J}\cos\phi$, $Y=\sqrt{2J}\sin\phi$, where $\phi$ is the reduced resonant angle formed from the two 2:1 critical arguments $2\lambda_d-\lambda_c-\varpi_c$ and $2\lambda_d-\lambda_c-\varpi_d$.  The parameter $\Gamma$ measures detuning from exact resonance and varies slowly under the prescribed migration.  For $\Gamma>3/2$, Equation~(\ref{eq:hamiltonian}) contains a stable resonant center, a saddle point, and a separatrix.  The separatrix encloses the capture region in which trajectories remain around the resonant center \citep{HenrardLemaitre1983}.  Residence inside this island and winding about the coordinate origin are distinct tests in the shifted $(X,Y)$ portrait; an apparent reversal or bounded interval of a critical angle therefore does not by itself establish island residence.  When $\Gamma\leq3/2$, the saddle and its enclosed island do not exist.

We store $a$, $e$, $\lambda$, and $\varpi$ for c and d every 0.05~yr from the first integration step through 20~yr after migration stops.  Before the canonical projection, we remove orbital-period oscillations with a 1~yr sliding quadratic fit (a Savitzky--Golay filter).  At each epoch for which an island exists, we calculate the trajectory energy $\mathcal H$ and the instantaneous saddle energy $\mathcal H_{\rm sep}$.  We define the dimensionless separatrix margin as
\begin{equation}
m=\frac{\mathcal H-\mathcal H_{\rm sep}}{1+|\mathcal H_{\rm sep}|}.
\end{equation}
Thus $m>0$ lies inside the island and $m<0$ lies outside it.  For $\Gamma\leq3/2$, $\mathcal H_{\rm sep}$ and $m$ are undefined; those epochs are recorded as having no resonant island, and no exterior margin is assigned.  The two $P_b/20$ reruns give 95th-percentile absolute margin differences of 0.0059 and 0.0087 relative to their $P_b/10$ parents, so the adopted $|m|=0.01$ operational buffer is slightly larger than the measured numerical scale.

The values 2~yr of island residence, 0.8 cycle of accumulated path about the moving resonant center, and the 0.01 margin buffer are operational thresholds for a secure entry classification.  A confirmed escape then requires a previously established island residence followed either by at least 1~yr robustly at $m<-0.01$, or by persistence for at least 1~yr after the tracked island disappears at $\Gamma\leq3/2$.  Resumed circulation of a critical angle is retained as auxiliary evidence.  Starting inside at the first stored epoch is labeled ``initially locked'' because the imposed initial phase placed the system in the island before migration.  Shorter or shallower excursions are labeled uncertain.  Repeating the census for residence thresholds of 1, 2, and 4~yr and path thresholds of 0.5, 0.8, and 1 cycle leaves zero entered-and-escaped trials throughout, except for the joint 1~yr--0.5 cycle choice, which admits the two short borderline paths described in Section~\ref{subsec:resonance}.  Dissipative capture can also become overstable and escape \citep{GoldreichSchlichting2014}, which motivates following the topology for the full passage.

Trajectories that remain outside receive a second test.  We integrate the same averaged c--d equations twice, retaining the secular background and imposed migration in both calculations.  One includes the two 2:1 harmonics and the other removes them.  We measure the detrended period-ratio anomaly and the change in the resonant eccentricity mode within the same window, with the rms far from 2:1 defining the background.  A run is ``outside, perturbed'' when the resonant calculation reproduces the N-body anomaly above three times that background and the calculation with those terms removed remains at the background level.

\begin{figure*}[t]
\centering
\includegraphics[width=0.97\textwidth]{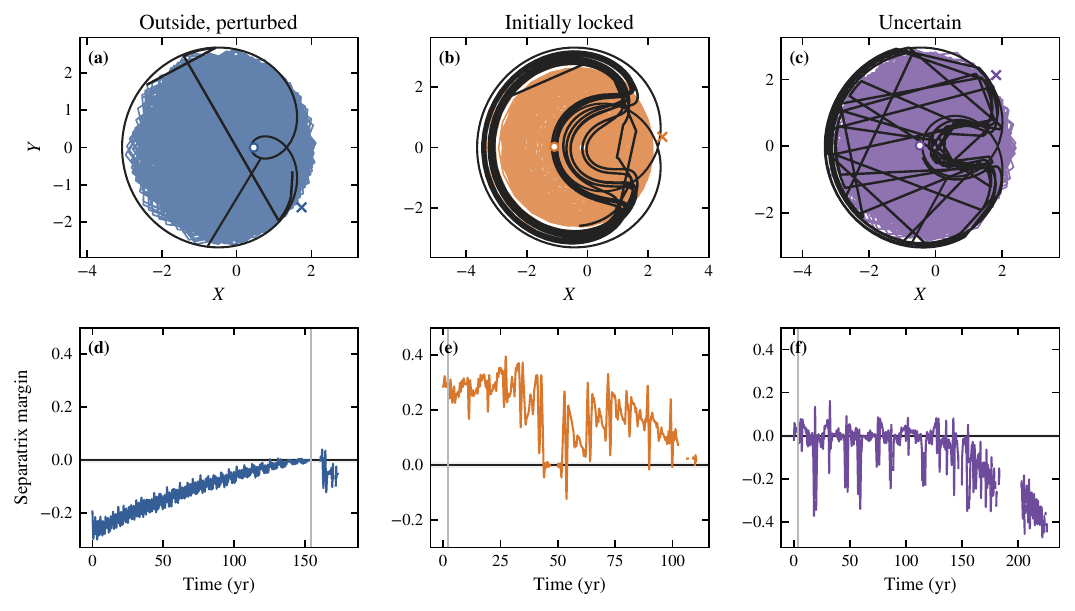}
\caption{Representative classifications from the forced c--d 2:1 experiment.  Columns show an exterior resonantly perturbed passage, a trajectory initialized inside the resonant island that later escapes, and an uncertain separatrix-grazing passage.  Top panels project the smoothed N-body paths into the canonical $(X,Y)$ plane; black curves are the instantaneous separatrices, open circles mark the first saved state, and crosses mark the final state.  Bottom panels show the normalized separatrix margin versus time, with positive values inside the island.  The first example remains exterior even though the 2:1 terms measurably alter its eccentricity vector and period-ratio drift.  The second begins locked before migration and therefore supplies no capture event.  The third enters only briefly and fails the duration or angular-path criterion.}\label{fig:resonance_diagnostics}
\end{figure*}

Among the 60 independent trials, the census contains zero entered-and-escaped trajectories, 20 outside-but-perturbed passages, nine initially locked trajectories that escape, and 31 uncertain passages.  No independent run remains locked at the end, and every securely exterior run shows a measurable perturbation.  Runs 61 and 62 are the two half-step convergence calculations, paired with independent runs 20 and 21; both reruns and both parents remain uncertain and are not counted again.  Figure~\ref{fig:resonance_diagnostics} shows how the phase-space criterion separates these outcomes.  The classification establishes that the 2:1 neighborhood can transfer eccentricity into b without sustained capture.  The circularization calculation in Section~\ref{subsec:resonance} supplies the independent timing constraint: c tides alone do not move the pair to the observed architecture before the injected b mode decays.

\bibliographystyle{aasjournalv7}
\bibliography{references}

@article{Kostov2019,
  author = {Kostov, Veselin B. and others},
  title = {The {L 98-59} System: Three Transiting, Terrestrial-size Planets Orbiting a Nearby {M} Dwarf},
  journal = {\aj},
  year = {2019},
  volume = {158},
  number = {1},
  pages = {32},
  doi = {10.3847/1538-3881/ab2459}
}

@article{Seligman2024,
  author = {Seligman, Darryl Z. and Feinstein, Adina D. and Lai, Dong and Welbanks, Luis and Taylor, Aster G. and Becker, Juliette and Adams, Fred C. and Morgan, Marvin and Bergner, Jennifer B.},
  title = {Potential Melting of Extrasolar Planets by Tidal Dissipation},
  journal = {\apj},
  year = {2024},
  volume = {961},
  number = {1},
  pages = {22},
  doi = {10.3847/1538-4357/ad0b82},
  eprint = {2311.01187},
  archiveprefix = {arXiv},
  primaryclass = {astro-ph.EP}
}

@article{BelloArufe2025,
  author = {Bello-Arufe, Aaron and Damiano, Mario and Bennett, Katherine A. and Hu, Renyu and Welbanks, Luis and MacDonald, Ryan J. and Seligman, Darryl Z. and Sing, David K. and Tokadjian, Armen and Oza, Apurva V. and Yang, Jeehyun},
  title = {Evidence for a Volcanic Atmosphere on the Sub-Earth {L 98-59 b}},
  journal = {\apjl},
  year = {2025},
  volume = {980},
  number = {2},
  pages = {L26},
  doi = {10.3847/2041-8213/adaf22},
  eprint = {2501.18680},
  archiveprefix = {arXiv},
  primaryclass = {astro-ph.EP}
}

@article{Nicholls2025,
  author = {Nicholls, Harrison and Guimond, Claire Marie and Hay, Hamish C. F. C. and Chatterjee, Richard D. and Lichtenberg, Tim and Pierrehumbert, Raymond T.},
  title = {Self-limited Tidal Heating and Prolonged Magma Oceans in the {L 98-59} System},
  journal = {\mnras},
  year = {2025},
  volume = {541},
  number = {3},
  pages = {2566--2584},
  doi = {10.1093/mnras/staf1167},
  eprint = {2505.03604},
  archiveprefix = {arXiv},
  primaryclass = {astro-ph.EP}
}

@article{Cadieux2025,
  author = {Cadieux, Charles and L'Heureux, Alexandrine and Piaulet-Ghorayeb, Caroline and Doyon, Ren{\'e} and Artigau, {\'E}tienne and Cook, Neil J. and Coulombe, Louis-Philippe and Roy, Pierre-Alexis and Lafreni{\`e}re, David and Lamontagne, Pierrot and Radica, Michael and Benneke, Bj{\"o}rn and Ahrer, Eva-Maria and Weisserman, Drew and Cloutier, Ryan},
  title = {Detailed Architecture of the {L 98-59} System and Confirmation of a Fifth Planet in the Habitable Zone},
  journal = {\aj},
  year = {2025},
  volume = {170},
  number = {3},
  pages = {154},
  doi = {10.3847/1538-3881/adef59},
  eprint = {2507.09343},
  archiveprefix = {arXiv},
  primaryclass = {astro-ph.EP}
}

@article{FarhatChiang2026,
  author = {Farhat, Mohammad and Chiang, Eugene},
  title = {Magma Ocean Waves and Thermal Variability on Lava Worlds},
  journal = {\apj},
  year = {2026},
  volume = {1003},
  number = {2},
  pages = {208},
  doi = {10.3847/1538-4357/ae6504},
  eprint = {2601.07080},
  archiveprefix = {arXiv},
  primaryclass = {astro-ph.EP}
}

@article{EngleGuinan2023,
  author = {Engle, Scott G. and Guinan, Edward F.},
  title = {Living with a Red Dwarf: The Rotation--Age Relationships of {M} Dwarfs},
  journal = {\apjl},
  year = {2023},
  volume = {954},
  number = {2},
  pages = {L50},
  doi = {10.3847/2041-8213/acf472}
}

@article{Mardling2007,
  author = {Mardling, Rosemary A.},
  title = {Long-term Tidal Evolution of Short-period Planets with Companions},
  journal = {\mnras},
  year = {2007},
  volume = {382},
  number = {4},
  pages = {1768--1790},
  doi = {10.1111/j.1365-2966.2007.12500.x}
}

@article{HansenMurray2015,
  author = {Hansen, Bradley M. S. and Murray, Norm},
  title = {Secular Effects of Tidal Damping in Compact Planetary Systems},
  journal = {\mnras},
  year = {2015},
  volume = {448},
  number = {2},
  pages = {1044--1059},
  doi = {10.1093/mnras/stv049}
}

@article{DriscollBarnes2015,
  author = {Driscoll, Peter E. and Barnes, Rory},
  title = {Tidal Heating of Earth-like Exoplanets around {M} Stars: Thermal, Magnetic, and Orbital Evolutions},
  journal = {Astrobiology},
  year = {2015},
  volume = {15},
  number = {9},
  pages = {739--760},
  doi = {10.1089/ast.2015.1325}
}

@article{Kervazo2021,
  author = {Kervazo, Mathilde and Tobie, Gabriel and Choblet, Ga{"e}l and Dumoulin, Caroline and B{\v{e}}hounkov{\'a}, Marie},
  title = {Solid Tides in {Io}'s Partially Molten Interior: Contribution of Bulk Dissipation},
  journal = {\aap},
  year = {2021},
  volume = {650},
  pages = {A72},
  doi = {10.1051/0004-6361/202039433}
}

@article{Peale1979,
  author = {Peale, Stanton J. and Cassen, Patrick and Reynolds, Ray T.},
  title = {Melting of {Io} by Tidal Dissipation},
  journal = {Science},
  year = {1979},
  volume = {203},
  number = {4383},
  pages = {892--894},
  doi = {10.1126/science.203.4383.892}
}

@article{Deck2013,
  author = {Deck, Katherine M. and Payne, Matthew and Holman, Matthew J.},
  title = {First-order Resonance Overlap and the Stability of Close Two-planet Systems},
  journal = {\apj},
  year = {2013},
  volume = {774},
  number = {2},
  pages = {129},
  doi = {10.1088/0004-637X/774/2/129}
}

@book{MurrayDermott1999,
  author = {Murray, Carl D. and Dermott, Stanley F.},
  title = {Solar System Dynamics},
  publisher = {Cambridge University Press},
  address = {Cambridge},
  year = {1999},
  isbn = {978-0-521-57295-8}
}

@article{ReinLiu2012,
  author = {Rein, Hanno and Liu, S.-F.},
  title = {{REBOUND}: An Open-Source Multi-Purpose {$N$}-Body Code for Collisional Dynamics},
  journal = {\aap},
  year = {2012},
  volume = {537},
  pages = {A128},
  doi = {10.1051/0004-6361/201118085}
}

@article{ReinTamayo2015,
  author = {Rein, Hanno and Tamayo, Daniel},
  title = {{WHFast}: A Fast and Unbiased Implementation of a Symplectic {Wisdom--Holman} Integrator for Long-term Gravitational Simulations},
  journal = {\mnras},
  year = {2015},
  volume = {452},
  number = {1},
  pages = {376--388},
  doi = {10.1093/mnras/stv1257}
}

@article{Tamayo2020,
  author = {Tamayo, Daniel and Rein, Hanno and Shi, Pengshuai and Hernandez, David M.},
  title = {{REBOUNDx}: A Library for Adding Conservative and Dissipative Forces to Otherwise Symplectic {$N$}-Body Integrations},
  journal = {\mnras},
  year = {2020},
  volume = {491},
  number = {2},
  pages = {2885--2901},
  doi = {10.1093/mnras/stz2870}
}

@article{Harris2020,
  author = {Harris, Charles R. and others},
  title = {Array Programming with {NumPy}},
  journal = {Nature},
  year = {2020},
  volume = {585},
  number = {7825},
  pages = {357--362},
  doi = {10.1038/s41586-020-2649-2}
}

@article{Virtanen2020,
  author = {Virtanen, Pauli and others},
  title = {{SciPy} 1.0: Fundamental Algorithms for Scientific Computing in {Python}},
  journal = {Nature Methods},
  year = {2020},
  volume = {17},
  number = {3},
  pages = {261--272},
  doi = {10.1038/s41592-019-0686-2}
}

@article{Hunter2007,
  author = {Hunter, John D.},
  title = {{Matplotlib}: A {2D} Graphics Environment},
  journal = {Computing in Science \& Engineering},
  year = {2007},
  volume = {9},
  number = {3},
  pages = {90--95},
  doi = {10.1109/MCSE.2007.55}
}

@article{GreenbergVanLaerhoven2011,
  author = {Greenberg, Richard and Van Laerhoven, Christa},
  title = {Tidal Evolution of a Secularly Interacting Planetary System},
  journal = {\apj},
  year = {2011},
  volume = {733},
  number = {1},
  pages = {8},
  doi = {10.1088/0004-637X/733/1/8},
  eprint = {1103.2794},
  archiveprefix = {arXiv},
  primaryclass = {astro-ph.EP}
}

@article{VanEylenAlbrecht2015,
  author = {Van Eylen, Vincent and Albrecht, Simon},
  title = {Eccentricity from Transit Photometry: Small Planets in {Kepler} Multi-planet Systems Have Low Eccentricities},
  journal = {\apj},
  year = {2015},
  volume = {808},
  number = {2},
  pages = {126},
  doi = {10.1088/0004-637X/808/2/126},
  eprint = {1505.02814},
  archiveprefix = {arXiv},
  primaryclass = {astro-ph.EP}
}

@article{Peterson2023,
  author = {Peterson, Merrin S. and Benneke, Bj{\"o}rn and Collins, Karen and others},
  title = {A Temperate Earth-sized Planet with Tidal Heating Transiting an {M6} Star},
  journal = {Nature},
  year = {2023},
  volume = {617},
  number = {7962},
  pages = {701--705},
  doi = {10.1038/s41586-023-05934-8}
}

@article{GreklekMcKeon2025,
  author = {Greklek-McKeon, Michael and Knutson, Heather A. and Levine, W. Garrett and Hu, Renyu and Saidel, Morgan and Gomez Barrientos, Jonathan and Batygin, Konstantin and Benneke, Bj{\"o}rn},
  title = {Updated Mass, Eccentricity, and Tidal Heating Constraints for the Earth-sized Planet {LP 791-18 d}},
  journal = {\aj},
  year = {2025},
  volume = {170},
  number = {2},
  pages = {65},
  doi = {10.3847/1538-3881/addf2d},
  eprint = {2501.18700},
  archiveprefix = {arXiv},
  primaryclass = {astro-ph.EP}
}

@article{Hu2024,
  author = {Hu, Renyu and Bello-Arufe, Aaron and Zhang, Michael and others},
  title = {A Secondary Atmosphere on the Rocky Exoplanet {55 Cancri e}},
  journal = {Nature},
  year = {2024},
  volume = {630},
  number = {8017},
  pages = {609--612},
  doi = {10.1038/s41586-024-07432-x},
  eprint = {2405.04744},
  archiveprefix = {arXiv},
  primaryclass = {astro-ph.EP}
}

@article{Barr2018,
  author = {Barr, Amy C. and Dobos, Vera and Kiss, L{\'a}szl{\'o} L.},
  title = {Interior Structures and Tidal Heating in the {TRAPPIST-1} Planets},
  journal = {\aap},
  year = {2018},
  volume = {613},
  pages = {A37},
  doi = {10.1051/0004-6361/201731992},
  eprint = {1712.05641},
  archiveprefix = {arXiv},
  primaryclass = {astro-ph.EP}
}

@article{Lainey2009,
  author = {Lainey, Val{\'e}ry and Arlot, Jean-Eudes and Karatekin, {\"O}zg{\"u}r and Van Hoolst, Tim},
  title = {Strong Tidal Dissipation in {Io} and {Jupiter} from Astrometric Observations},
  journal = {Nature},
  year = {2009},
  volume = {459},
  number = {7249},
  pages = {957--959},
  doi = {10.1038/nature08108}
}

@article{Segatz1988,
  author = {Segatz, M. and Spohn, T. and Ross, M. N. and Schubert, G.},
  title = {Tidal Dissipation, Surface Heat Flow, and Figure of Viscoelastic Models of {Io}},
  journal = {Icarus},
  year = {1988},
  volume = {75},
  number = {2},
  pages = {187--206},
  doi = {10.1016/0019-1035(88)90001-2}
}

@article{HenrardLemaitre1983,
  author = {Henrard, Jacques and Lema{\^i}tre, Anne},
  title = {A Second Fundamental Model for Resonance},
  journal = {Celestial Mechanics},
  year = {1983},
  volume = {30},
  number = {2},
  pages = {197--218},
  doi = {10.1007/BF01234306}
}

@article{MustillWyatt2011,
  author = {Mustill, Alexander J. and Wyatt, Mark C.},
  title = {A General Model of Resonance Capture in Planetary Systems: First- and Second-order Resonances},
  journal = {\mnras},
  year = {2011},
  volume = {413},
  number = {1},
  pages = {554--572},
  doi = {10.1111/j.1365-2966.2011.18201.x},
  eprint = {1011.4804},
  archiveprefix = {arXiv},
  primaryclass = {astro-ph.EP}
}

@article{GoldreichSchlichting2014,
  author = {Goldreich, Peter and Schlichting, Hilke E.},
  title = {Overstable Librations Can Account for the Paucity of Mean-motion Resonances among Exoplanet Pairs},
  journal = {\aj},
  year = {2014},
  volume = {147},
  number = {2},
  pages = {32},
  doi = {10.1088/0004-6256/147/2/32},
  eprint = {1308.4688},
  archiveprefix = {arXiv},
  primaryclass = {astro-ph.EP}
}

\end{document}